\documentclass[amsmath,amssymb,pre,twocolumn,showpacs,floatfix]{revtex4}

\newcommand{\be}{\begin{equation}}
\newcommand{\ee}{\end{equation}}
\newcommand{\ba}{\begin{eqnarray}}
\newcommand{\ea}{\end{eqnarray}}
\newcommand{\rd}{\mathrm{d}}            
\newcommand{\re}{\mathrm{e}}            

\newcommand{\te}[1]{\mbox{\boldmath $ #1 $}}

\newcommand{\bx}{\te{x}}

\newcommand{\bbe}{\te{e}}
\newcommand{\bu}{\te{u}}

\newcommand\eg{\textit{e.g.}\ }
\newcommand\ie{\textit{i.e.}\ }

\usepackage{graphicx}
\usepackage{dcolumn}
\usepackage{times}

\begin{document}

\title{Do small swimmers mix the ocean?}
\author {A.\ M.\ Leshansky$^{1}$}\email{lisha@technion.ac.il}
\author{L.\ M.\ Pismen$^{1,2}$}
\affiliation{$^1$Department of Chemical Engineering, $^2$Minerva Center for Nonlinear Physics of Complex Systems, Technion -- Israel Institute of Technology, Haifa 32000, Israel}

\date{\today}

\begin{abstract}
In this communication we address some hydrodynamic aspects of recently proposed drift mechanism of biogenic mixing in the ocean [K.~Katija and  J.~O. Dabiri, Nature \textbf{460}, 624 (2009)]. The relevance of  the locomotion gait at various spatial scales with respect to the drift is discussed.  A hydrodynamic scenario of the drift based on unsteady inertial propulsion, typical for most small marine organisms, is proposed. We estimate its effectiveness by taking into account interaction of a swimmer with the turbulent marine environment. Simple scaling arguments are derived to estimate the comparative role of drift-powered mixing with respect to the turbulence. The analysis indicates substantial biomixing effected by relatively small but numerous drifters, such as krill or jellyfish.
\end{abstract}

\pacs{47.63.M-, 92.10.-c}

\maketitle

An intriguing question vigorously discussed in recent literature is whether marine organisms contribute significantly to the ocean mixing. The arguments both in favor \cite{biomix06, biomix06a} and against \cite{antimix07} a substantial role of biomixing have been based on the overall energy balance, and assumed that a certain part of the overall energy intake of marine organisms contributes to mixing rather than being dissipated by other means. This approach may be inadequate in view of uncertainties in estimating alternative dissipation pathways without considering swimming dynamics explicitly. Very recent work \cite{dabiri09} suggests that the net displacement of fluid particles by swimmers (known as ``drift mechanism" \cite{darwin53}), being enhanced by viscosity, is more relevant for mixing than biologically generated turbulence. Experimental observation of drift behind swimming jellyfish and hydrodynamic demonstration estimating drift capacity of passively towed objects at finite Reynolds numbers, Re, \cite{dabiri09} give support to the notion of strong biomixing. It is the aim of this communication (i) to examine the hydrodynamics of the drift induced by  self-propelled objects; (ii) to derive mechanistic estimates based on scaling arguments, taking into account the turbulent environment and comparing swimmer's contribution to mixing with turbulent stirring on relevant scales.

The major factors determining the answer to the question in the title are the \emph{drift volume} $D$ dragged by the swimmer and the \emph{turbulent scales} characteristic of the marine environment. Both factors may vary within a wide range.

A steadily towed body in a viscous fluid drags an infinite volume of fluid when the flow is dominated by viscosity \cite{EGD03}, while in inviscid potential flow it is of the same order of magnitude as the volume of the object \cite{darwin53}. Recent demonstration of the drift due to a passively towed two-dimensional body of the dimension $\ell_b$ at $\mathrm{Re}=U \ell_b/\nu \sim 5$--$100$ in \cite{dabiri09} indicates that the viscosity-enhanced drift volume may far exceed the swimmer's own volume. This tendency can be readily explained. Advection of vorticity, shed by a body towed steadily with the speed $U$ through a fluid quiescent at infinity, is balanced at some distance downstream by transverse viscous diffusion. The asymptotic solution for the velocity deficiency in the wake $v=U-u$ (in the coordinate frame comoving with the swimmer in the negative $x$ direction) can be derived assuming small departure from the free stream velocity at positions sufficiently far downstream, $v \ll U$, and using the boundary layer approximation, $\partial/\partial x \ll \partial/\partial r$, where $r$ is the transverse radial distance. The asymptotic solution at $x \rightarrow \infty$ is $v \sim (J/4\pi\nu x)\: \exp{(-U r^2/4\nu x)}$ \cite{batchelor67}. Here $\te{J} \delta(\bx)$ is the point force yielding the kinematic momentum flux (force per unit mass) $\te{J}=-J \bbe$, where $\bbe$ is the unit vector along the $x$-axis. The flow disturbance outside the wake away from the object is irrotational radial flow due to ``potential source", $\bu=(J/4\pi U) \: \bx/|\bx|^3$, compensating for the mass deficiency in the wake \cite{batchelor67}. The rate at which the fluid volume is dragged across a stationary plane behind the swimmer, $q=2\pi\int_{r=0}^{\infty} v  r\rd r=J/U$, can be related to the net drag force exerted on the body, $F=\rho J$, where $\rho$ is the fluid density \cite{batchelor67}. Clearly, since $J=$ const, the corresponding drift volume $D$ is infinite, similar to the zero Re case \cite{EGD03}. Considering the partial drift volume $D_p$ dragged by a body that had been towed a distance $L$ over time $L/U$ \cite{dabiri09}, and expressing the drag as $F=\frac{1}{2} C_D \rho\: U^2 \mathcal{A}$, where $\mathcal{A}$ is the wetted surface area and $C_D$ is the drag coefficient, we find that $D_p=q L/U=\frac{1}{2}\: C_D \mathcal{A} L$. Since $C_D$ is a decreasing function of Re, such that $C_D \sim$ Re$^{-\alpha}$ with the exponent $\alpha=1$ at Re$\lesssim 1$, and varying from $1/2$ to $1/5$, respectively, in the laminar and turbulent boundary layer regime at high Re, the concept of viscosity-enhanced drift behind passively towed body demonstrated numerically in \cite{dabiri09} becomes clear, as lower Re results in the increased partial drift volume. The question is: to what extent this notion is relevant towards drift induced by self-propelled organisms?

The flow field around self-propelled objects could be, however, quite different from that of a passively towed body. A \emph{steadily} self-propelled swimmer generates no net momentum flux, since the thrust is counter-balanced by the drag force \cite{wakes,SV03}. The two forces of equal magnitude, acting in opposite directions and separated by some distance, constitute a force dipole, that can be approximated away from the body as $Q_{ij} \nabla_j \delta(\bx)$  \cite{SV03}. Higher order terms, such as force quadrupole, $M_{ijk} \nabla_j \nabla _k \delta(\bx)$, etc., are decaying faster. The net momentum flux induced by such force dipole is zero and therefore the corresponding drift volume is expected to be far smaller than for a passively towed body.

The drift induced by a high-Re steadily propelled swimmer can be estimated along the same lines as that by a passively towed object considered above. The far-field velocity deficiency in the zero-momentum wake behind a localized force dipole in a spatially uniform stream of velocity $U$ along the $x$-axis is given by $v=(Q/4\pi\nu x^2)\: (1-\eta)\: \re^{-\eta}$ at $x \rightarrow \infty$, where $\eta=U r^2/4\nu x$ \cite{SV03}. The far flow field outside the wake is due to a potential doublet, $\bu=(-Q/4\pi U |\bx|^3) \{\bbe -3 (\bbe\cdot \bx)\, \bx/|\bx|^2\}$, and is the same as the far field of a steadily moving object in potential flow \cite{batchelor67}. It can be readily shown that the net fluid flux $q$ across a stationary plane in the wake is zero. Whereas for a monopole forcing exerted on a passively towed object the radial potential flow outside the wake is compensated by a net influx in the wake, for a dipole forcing the potential doublet-like flow outside the wake causes no net fluid drift, and, as result, there is no net drift within the wake as well. Thus, the net drift volume will be comparable to the swimmer's volume as in potential flow.

Although small swimmers ($<$1 mm wide) are mainly nonvertical migrators \cite{Hays94}, it is instructive to examine the drift associated with their motion. In the viscous regime, the mean far-field flow (axisymmetric with respect to $x$-axis) around a force-free swimmer propelled with the mean velocity $U$ in the negative $x$-direction through unbounded fluid quiescent at infinity, can be written in the comoving frame as
\be
\psi \approx \left(-\frac{1}{2} U r^2 +\frac{M}{\nu r}\right)\sin^2{\theta}+\frac{Q}{\nu} \cos{\theta}\sin^2{\theta}, \label{eq:psi1}
\ee
where $Q$ and  $M$ are, respectively, the strengths of the force dipole and quadrupole, and $r,\:\theta$ are spherical coordinates with $\theta$ measured from the positive $x$ axis. A swimmer with negative $Q$ is called a ``puller", as for micro-organisms having their thrust-generating apparatus in front of the body (which dominates the drag), \eg biflagellate algae such as \emph{Chlamydomonas}. An organism with positive $Q$ is a ``pusher", i.e. the thrust is generated behind the body, as in bacteria or spermatozoa. Proceeding along the same lines as in Ref.~\cite{EGD03}, it is convenient to express the radius of the streamline tube $\xi_0$ far up or downstream from the swimmer (which are equal due to fore-and-aft symmetry), through $\psi=-U \xi_0^2/2$. Expanding for large $\xi_0$ and solving for $r$ yields:
\be \label{eq:rinf}
r=\frac{\xi_0}{\sin{\theta}} \left\{1+\frac{Q \cos{\theta} \sin^2{\theta}}{\xi_0^2\:U}
+\frac{M \sin^3{\theta}}{\xi_0^3\: U}+\mathcal{O}(\xi_0^{-4})\right\}.
\ee
The drift volume $D$ is defined as the fluid volume between an undisturbed material plane which is initially far in front of the swimmer, and the final position of the same surface disturbed by the swimmer moving off to infinity \cite{darwin53}. The partial drift volume can be defined for a material surface of a finite radius $\xi_0$ \cite{EGD03}, $D_p=2\pi\int_0^{\xi_0}  X \xi_0 \rd \xi_0$, where $X$ is the displacement of a fluid particle in the direction of translation. Taking the limit of $D_p$ at $\xi_0 \rightarrow \infty$ results in the drift volume $D$. Using the integral mass conservation arguments \cite{EGD03}, the partial drift volume can be found as $D_p=A-V$, where $V$ is the swimmer's volume and $A$ is the volume between the laterally displaced streamline tube and its unperturbed position, $A=\pi \int_{x_o}^{x_i}  \left(r^2 \sin{\theta}^2-\xi_0^2 \right) \rd x$, with $x_i$ and $x_o$ being the coordinates of the streamline tube inlet (upstream) and outlet (downstream). Using $r(\xi_0)$ in Eq.~(\ref{eq:rinf}), we change the integration variable to $\theta$ via $x=\xi_0 \cot{\theta}$ and evaluate the integral as $A=4M\pi/U\nu+\mathcal{O}(\xi_0^{-3})$ at $\xi_0 \rightarrow \infty$. Thus, the drift volume for a force-free viscous swimmer is finite,
\be
D = 4M\pi/U\nu-V\:.
\ee
Note that the drift volume is independent of $Q$ whereas both ``pushers" and ``pullers" would displace the same amount of fluid. The value of $M$ and the corresponding drift volume depends on particular swimming gait. For instance, it can be evaluated for a model ``spherical squirmer" \cite{squirmer} of radius $a$ propelled by purely tangential axisymmetric and time-independent surface motion, that in the co-moving reference frame can be written as $\bu|_{r=a}=u_s (1+\beta \cos{\theta}) \sin{\theta} \bbe_{\theta}$, where $u_s$ is the magnitude of the surface velocity. The propulsion speed and dipole and quadrupole forcing intensity in (\ref{eq:psi1}) can be readily found upon application of the latter boundary condition at $r=a$, yielding $U=2 u_s/3 $, $M=a^3 u_s \nu/3$, $Q=-a^2 \beta u_s \nu/2$. Therefore, the drift volume for a ``spherical squirmer" is finite and equal to half of its volume, $D=4M\pi/U\nu-V=V/2$. The same result was found for a rigid sphere moving through an inviscid fluid \cite{darwin53}.

Although the above analysis shows no considerable drift and the anticipated drift volume is comparable to that found for inviscid potential flow, yet a substantial drift observed in experiments with dye dispersion behind a jellyfish  \cite{dabiri09} can be attributed to \emph{unsteady} locomotion, typical for small \emph{inertial} marine swimmers \cite{vogel}. For an organism that swims unsteadily, such as jellyfish propelled by jetting, or small crustaceans propelled by rowing their legs, the instantaneous thrust must be balanced by the force to overcome its inertia as well as two hydrodynamic forces: drag and the acceleration reaction (or the added mass force) \cite{vogel,daniel,McHJ03}. During a fast propulsion stage (power stroke), the swimmer transfers momentum to the surrounding fluid generating accelerated propulsion, as the thrust exceeds the drag force and the reactive force by the amount equal to the swimmer's inertia.  During a slow retraction stage, the swimmer decelerates absorbing back some momentum, but the overall net momentum flux over a stroke cycle transferred to the fluid is non-zero due to a nonlinear nature of the inertial propulsion. This may yield a considerable drift due to a momentum/mass deficit in the wake in the same way as for a passively towed object discussed above. These arguments may shed light on why numerical demonstration of viscosity-enhanced drift behind a passively towed object, provided in \cite{dabiri09} without a comment or justification, is relevant, to some extent, to marine self-locomotion.

In turbulent environment, the drift volume is not infinite, as it is dispersed by the background turbulence. The turbulent marine environment is characterized by the Kolmogorov scale $\ell_k=(\nu^3/\varepsilon)^{1/4}$, where $\nu$ is the kinematic viscosity and $\varepsilon$ is the dissipation rate per unit mass of the fluid, and the Ozmidov scale, $\ell_{o}=(\varepsilon/N^3)^{1/2}$, corresponding to the largest eddy that can overturn for prescribed stable density stratification in the ocean $\rd \rho/\rd z$. Here  $N=\sqrt{-(g/\rho_0)(\rd\rho/\rd z)}$ is the buoyancy frequency, with $\rho_0$ being the mean fluid density \cite{thorpe05}.

When the swimmer is propelled though a weakly turbulent fluid, the momentum wake fades away at a distance $\ell$ downstream where the velocity deficiency $v$ becomes comparable to the magnitude of turbulent velocity fluctuation $u'$. Moreover, the vorticity in the wake is diffused transversely by turbulent eddies with the the scale-dependent turbulent dispersion coefficient, $\mathcal{K}$. Although the detailed analysis of the wake in unsteady locomotion in a weakly turbulent environment is beyond the scope of the present study, a rough estimate of the mean momentum flux yields $J=(m/\rho) (\Delta U/\Delta \tau)$, where $\Delta U$ is the velocity gain by a swimmer of mass $m$ during a pulse phase of duration $\Delta \tau$. Taking constant eddy diffusivity corresponding to the largest eddy, and assuming that the time-averaged structure of the momentum wake is the same as for a passively towed body, the comparison between the velocity deficit in the momentum wake responsible for the drift, $v\sim J/\mathcal{K} \ell $, and $u'$ yields the following scaling estimate for the apparent wake length, $\ell$,
\be
\label{wake}
\ell\sim  (m \Delta U)/(\Delta \tau \rho \mathcal{K} u')\:. 
\ee
For instance, the jellyfish \emph{Aurelia aurita} with the bell diameter of $\sim 8.5$ cm and mass $m \sim 20$ gr, gains $\Delta U \sim 3$ cm~s$^{-1}$ during $\Delta \tau \sim 0.5$ s \cite{McHJ03}. For a typical turbulent patch in a seasonal thermocline with $\varepsilon=10^{-8}$ W~kg$^{-1}$ and buoyancy frequency $N \sim 0.01$ s$^{-1}$ \cite{YS96}, the turbulent velocity fluctuations at  the Ozmidov scale, $u'_o \sim (\varepsilon \ell_o)^{1/3}$ are about $1$ mm s$^{-1}$,  and the corresponding vertical eddy diffusivity (also known as density or diapycnal diffusivity, \cite{thorpe05}) $\mathcal{K}\sim u_o' \ell_o=\varepsilon/N^2 \sim 10^{-4}$ m$^2$~s$^{-1}$. Thus, the estimate (\ref{wake}) yields a surprising $\ell \sim 10$ m! However, the strength of the momentum flux diminishes rapidly with the organism dimensions as expected. For a smaller jellyfish with the bell diameter of $1.8$ cm and mass $0.5$~gr, which gains $\Delta U \sim 1.2$ cm~s$^{-1}$ in $\Delta \tau \sim 0.3$~s \cite{McHJ03}, the estimate (\ref{wake}) yields $\ell \sim 20$ cm only.

Note that the proposed mechanism leading to (\ref{wake}) does not offer any viscosity-enhancement of the drift. On the opposite, smaller swimmers are expected to be less efficient drifters. For small viscous swimmers (propelled either steadily or unsteadily) the proposed mechanism is not operative. However, tiny swimmers can be better drifters when they swim against an external force, such as gravity. In this case they generate a slowly decaying Stokeslet in the far field, analogously to a passively towed object \cite{EGD03}. This situation arises \eg during upward active swimming or downward passive sinking of bottom-heavy swimmers. The net momentum transferred to the fluid will be proportional in this case to the swimmer's excess weight and will be aligned with the direction of gravity regardless of the direction of their motion.

The volume of a typical size $\ell>\ell_k$ is dragged along by a swimmer, which can be considered moving independently of the weakly turbulent local flow \cite{YS96}, for a time $\tau$ which takes for turbulent pulsations \emph{on the same scale} $\ell $ to disperse it. The estimate for this time is $\tau \sim \ell/u'$, where $u'$ is the characteristic turbulent fluctuation velocity on the scale $\ell$. It can be estimated, assuming it obeys the isotopic Kolmogorov scaling in the inertial sub-range, as $u' \sim (\varepsilon \ell)^{1/3}$. The resulting effective dispersion coefficient is $\mathcal{D} \sim \phi\,U^2 \tau \sim \phi\, U^2 \ell/u'$, where $U$ is the swimmer velocity and $\phi$ is the volume fraction occupied by the swimmers. This can be compared with the vertical eddy diffusivity $\mathcal{K}\sim \varepsilon/N^2$, leading to an order-of-magnitude estimate
\be
\label{dka}
\mathcal{D}/\mathcal{K} \sim \phi\,U^2 N^2 \ell^{2/3}\varepsilon^{-4/3}\:.
\ee

For small viscous swimmers (Re$=U \ell_s/\nu \lesssim 1$) of a size $\ell_s \lesssim \ell_k$, with a swimming speed lower than the rms turbulence velocity, such as phytoplankton \cite{YS96}, the typical size of the dragged volume accessible to dispersion by turbulent pulsations is comparable to $\ell_k$. The typical dispersion time is defined in this case on the finest turbulence scale and is estimated as $\tau \sim (\nu/\varepsilon)^{1/2}$ leading to $\mathcal{D}\sim \phi\: U^2 (\nu/\varepsilon)^{1/2}$. This should be compared with the turbulent dispersion coefficient $\mathcal{K}$ to give
\be
\label{dk1a}
\mathcal{D}/\mathcal{K} \sim \phi\,U^2 N^2 \nu^{1/2} \varepsilon^{-3/2}\:.
\ee
Note that the latter estimate does not depend on the swimmer's size.

For large inertial swimmers corresponding to Re $\gg 1$, where the potential flow assumption is applicable, $\ell \sim \ell_s$ \cite{darwin53} and the dispersion time is restricted to $\tau \sim\ell_s/U$, as a tracer particle interacts with a swimmer for a typical time that takes a swimmer to pass by. This modifies the dispersion estimate to $\mathcal{D} \sim \phi\: U\ell_s$. This scaling-based estimate coincides, up to an $O(1)$ shape-dependent multiplicative constant, with the results \cite{childress} of rigorous analysis of stirring by 2D and 3D bodies propelled in the potential flow, which have shown that displacement of passive tracer particles is completely dominated by ``head-on" collisions with the bodies. Comparing to vertical dispersion by turbulent buoyancy flux yields
\be
\label{dk2a}
\mathcal{D}/\mathcal{K} \sim \phi\, U\,\ell_s\, N^2 \varepsilon^{-1}\:.
\ee
However, the assumption of potential flow does not take into account a wake shed by the self-propelled body and the prediction (\ref{dk2a}) may considerably underestimate the contribution of drift-induced mixing.

The above estimate of turbulent dispersion $\mathcal{K}\sim \varepsilon/N^2$ suggested by scaling is valid in the \emph{transitional} regime corresponding to strong density stratification, where the parameter $I=\varepsilon/\nu N^2$ (\ie the squared ratio of the buoyancy time scale $N^{-1}$ to the time $(\nu/\varepsilon)^{1/2}$ required for turbulent events to develop)  is in the range $7\lesssim\ I \lesssim 100$ \cite{vertmix08}. Beyond this range $\mathcal{K}\sim \varepsilon/N^2$ greatly overestimates the actual turbulent dispersion. In the \emph{energetic} regime corresponding to weak stratification, where $I \gtrsim 100$, turbulence is only weakly affected by buoyancy as it develops relatively fast, and the dispersion coefficient is  $\mathcal{K} \sim \nu (\varepsilon/\nu N^2)^{1/2}$ \cite{vertmix08}.  The use of the latter scaling estimate for $\mathcal{K}$ modifies the ratio $\mathcal{D}/\mathcal{K}$ in Eqs.~(\ref{dka})--(\ref{dk2a}) resulting in even stronger contribution of the drift to mixing in the energetic regime.

Let us estimate the relative effect of mixing caused by vertically migrating Antarctic krill \emph{Euphausia pacifica}. The buoyant frequency $N$ for a strongly stratified ocean surface layer is typically about $10^{-2}$ s$^{-1}$ and the background dissipation rate measured during daylight when krill remained stationary at depths $\sim$ 100 m is $\varepsilon \approx 10^{-9}$ W~kg$^{-1}$  \cite{biomix06}. This gives $I\sim 10$, and therefore (\ref{dka}) is applicable. As the detailed measurements of propulsion kinematics available for jellyfish (e.g. \cite{McHJ03}) have not been reported for krill, we cannot estimate $\ell$ using (\ref{wake}). We, therefore, take $\ell \approx \ell_s$ as a conservative estimate of the characteristic dragged volume dimension. The krill length $\ell_s \approx 1.5$--$2$ cm, the average swimming speed is $U \approx 2.5$--$3.5$ cm~s$^{-1}$, and volume concentration for a dense layer of krill is $\phi=0.07$--$0.14$\%, corresponding to $5,000$--$10,000$ individuals m$^{-3}$ \cite{biomix06}, whereas the volume occupied by a single krill is estimated as $\pi r^2 \ell_s$, with the krill radius is $r \approx 0.15$ cm \cite{krill03}. This yields $\mathcal{D}/\mathcal{K} \sim 5$--$15$ in Eq.~(\ref{dka}), which indicates a substantial biogenic contribution of the drift to the overall mixing.

In the seasonal thermocline stratification the buoyant frequency $N$ may exceed $0.05$ s$^{-1}$ at the depth between 40 and 240 m \cite{thorpe05}, so that turbulence is effectively suppressed, and biomixing is expected to be a major stirring mechanism. In this case, $I\sim 0.4$ corresponds to the \emph{molecular} regime, when $\mathcal{K}$ should be replaced in the comparison estimates by $\nu$ \cite{vertmix08}. In the latter case, $\mathcal{D}\sim \phi\: U^2 \tau$, where $\tau\sim\ell^2/\nu$.
Therefore, $\mathcal{D}/\nu \sim \phi\: U^2 \ell^2/\nu^2$. For Antarctic krill \cite{biomix06}, we have $\mathcal{D}/\nu \sim 50 $--$ 250$.  On the other hand, for the weakly stratified ocean, $N \sim 10^{-3}$ s$^{-1}$ and less, and assuming $\varepsilon \approx 10^{-9}$ W~kg$^{-1}$ we obtain $I\sim 1000$. This suggests that the ratio $D/\mathcal{K}$ should be estimated using $\mathcal{K} \sim \nu (\varepsilon/\nu N^2)^{1/2}$ corresponding to the energetic regime and yielding $\mathcal{D}/\mathcal{K} \sim 2$--$4$.

Most of the marine biomass is concentrated in small organisms, such as copepods ($\ell_s\sim1$-$3$ mm), with some of them exhibiting vertical migration, feeding near the surface at night, then sinking into deeper water during the day to avoid visual predators \cite{Hays94}. The question of whether such diel vertical migrations may contribute to significant dispersion enhancement is still under discussion (e.g. \cite{antimix07}). According to the estimates in Eq.~(\ref{dk1a}), small swimmers of the size $\ell_s<\ell_k$ may generate significant mixing only at relatively high densities, due to their small propulsion speed. Using typical copepod swimming speeds $u\sim 2 $--$ 5$ mm~s$^{-1}$ at a maximum volume concentration $1.5$--$ 2$\%  (based on the universal scaling for maximum packing in a swarm \cite{HZ04}) and the same values of $\varepsilon$ and $N$ as in Ref.~\cite{biomix06}, Eq.~(\ref{dk1a}) yields $\mathcal{D}/\mathcal{K}\sim 1$ in the transitional regime. In the energetic regime with $N=10^{-3}$ s$^{-1}$, Eq.~(\ref{dk1a}) gives $\mathcal{D}/\mathcal{K}\sim 0.1$. The mean population density of copepods in the ocean is estimated as $n<10^4$ m$^{-3}$ \cite{MCV09}, corresponding to $\phi<0.001$\%. For such low mean population density, the above estimates suggest that drift-induced contribution to mixing due to copepods vertical migration is negligible.

The drift may be, however, facilitated by potential \emph{collective} effects, as suggested in \cite{dabiri09}. The drift volumes of individual swimmers can be shared by many individuals within a swarm, so that coordinated migration may be responsible for the formation of a large collective trail on the swarm scale. Assuming regular spatial orientation of the drifters, one can estimate the mean nearest-neighbor distance in a swarm as $\sim 1.2\: n^{-1/3}$ \cite{HC79}. Using the universal scaling for maximum population density \cite{HZ04}, $n_m \sim 0.022\: \ell_s^{-3.6}$ one arrives at the mean nearest-neighbor distance in a dense swarm $\sim 4.25 \:\ell_s^{1.2}$. This estimate suggests that organisms with the body length $\sim$1 mm in a dense swarm are separated by about one body length on the average, while for organisms with $\ell_s\sim 1$ cm, the mean nearest-neighbor distance is about two body lengths. These estimates are in accord with population densities reported in the literature and calculated from photographic recording of aggregations, acoustic measurements and catches in nets and trawls. For the organisms in the size range $\ell_s \sim 1$--$3$ cm, such as \emph{Euphausia pacifica} 
and others, the observed nearest-neighbor distance in dense swarms is in the range of 2$-$4 body lengths (see Tab.~2 in \cite{ritz94}).

The intriguing problem of collective drift by hydrodynamically interacting self-propelled objects will be subject of further investigation.


\begin{thebibliography}{88}

\bibitem{biomix06} W. K. Dewar et al.,
J. Mar. Res. \textbf{64}, 541 (2006).

\bibitem{biomix06a} E. Kunze et al.,
Science \textbf{313}, 1768 (2006).

\bibitem{antimix07} A.~W. Visser,
Science \textbf{316}, 838 (2007).

\bibitem{dabiri09} K.~Katija and  J.~O. Dabiri,
Nature \textbf{460}, 624 (2009).

\bibitem{darwin53} C.~Darwin.
Proc. Camb. Phil. Soc. Biol. Sci. \textbf{49}, 342 (1953).

\bibitem{EGD03} I. Eames, D. Gobby  and S.~B. Dalziel.
J. Fluid Mech. \textbf{485}, 67 (2003).

\bibitem{batchelor67} G.~K. Batchelor, An Introduction to Fluid Dynamics (Cambridge Univ. Press, 1967).

\bibitem{wakes} G.~Birkhoff and E.~H. Zarantanello, Jets, Wakes, and Cavities (Academic Press, 1957); V.~L. Sennitski, J.~Appl.~Mech.~Tech.~Phys. \textbf{25}, 526 (1984); V.~V. Pukhnachev, ibid. \textbf{30}, 215 (1989); Y.~D. Afanasyev, Phys. Fluids \textbf{16}, 3235 (2004).

\bibitem{SV03} S.~A. Smirnov and S.~I. Voropayev,
Phys. Lett. A \textbf{307}, 148 (2003).

\bibitem{Hays94} G.~C. Hays et al. Limnol. Oceanogr. 39, 1621 (1994).

\bibitem{squirmer} J.~Lighthill, Comm. Pure Appl. Math. \textbf{5}, 109 (1952); J.~R. Blake J. Fluid Mech. \textbf{46}, 199 (1971); H. A. Stone and A.~D.~T. Samuel, Phys. Rev. Lett. \textbf{77}, 4102 (1996); T.~Ishikawa, J.~T. Locsei and T.~J. Pedley, J. Fluid Mech. \textbf{615}, 401 (2008).

\bibitem{vogel} S.~Vogel, Life in Moving Fluids: The Physical Biology of Flow (Princeton Univerisity Press, 1994).

\bibitem{daniel} T. L. Daniel
Can. J. Zool. \textbf{61}, 1406-1420 (1983); T.~L. Daniel,
J.~Exp.~Biol. 119, 149-164 (1985).

\bibitem{McHJ03} M.~J. McHenry and J. Jed,
J. Exp. Biol. 206, 4125 (2003).

\bibitem{thorpe05} S.~A. Thorpe, The Turbulent Ocean (Cambridge Univ. Press, 2005).

\bibitem{YS96} H. Yamazaki and K.~D. Squires,
Mar. Ecol. Prog. Ser. \textbf{144}, 299--301 (1996).

\bibitem{vertmix08} G.~N. Ivey, K.~B. Winters and J.~R. Koseff,
Ann. Rev. Fluid. Mech. \textbf{40}, 169 (2008).

\bibitem{childress} J.-L. Thiffeault and S. Childress,
arXiv:0911.5511v2.

\bibitem{HZ04} M.~E. Huntley and M.~Zhou,
Mar. Ecol. Prog. Ser. \textbf{273}, 65 (2004).

\bibitem{krill03} J.~Yen, J.~Brown and D.~R.~Webster,
Mar. Fresh. Behav. Physiol. \textbf{36}, 307 (2003).

\bibitem{MCV09} R.~Di Mauro, F. Capitanio and M. D. Vi\~nas,
Brazilian J. Oceanography, \textbf{57}, 205 (2009).

\bibitem{HC79} W.~M. Hamner and J.~H. Carleton,
Limnol. Oceanogr. \textbf{24}, 1 (1979).

\bibitem{ritz94} D.~A. Ritz,
Adv. Marine Biology \textbf{30}, 156 (1994).

\end{thebibliography}
\end{document}